\input harvmac.tex
\Title{\vbox{\baselineskip12pt\hbox{MRI-PHY/97-04}\hbox{hep-th/9702031}}}
{\vbox{\centerline{Non-perturbative Enhanced Gauge Symmetries in}
         \vskip2pt\centerline{the Gimon-Polchinski Orientifold}}}

\centerline{ Dileep P. Jatkar } 
\smallskip\centerline{\it Mehta Research Institute of Mathematics
and Mathematical Physics}
\centerline{\it 10, Kasturba Gandhi Marg, Allahabad 211 002, INDIA}

\vskip .3in
We study the enhanced gauge symmetries at constant couplings in the moduli 
space of the Gimon-Polchinski Orientifold. We show that whenever there is 
a singularity of $E_6$, $E_7$ and $E_8$ type we get enhanced gauge symmetry 
$U(5)$, $U(6)$ and $U(8)$ respectively. This is due to the additional twisting 
involved in the definition of the Gimon-Polchiniski orientifold.

\Date{02/97}

\lref\vafaI{C. Vafa, {\it Evidence for F-Theory}, hep-th/9602022.}

\lref\vafaII{C. Vafa and D. Morrison, {\it Compactifications of F-Theory on
Calabi-Yau Threefolds-I and -II}, hep-th/9602114, hep-th/9603161.}

\lref\senI{A. Sen, {\it F-Theory and Orientifolds}, hep-th/9605150.}

\lref\mukhi{K. Dasgupta and S. Mukhi, {\it F-Theory at Constant Coupling}, 
Phys. Lett. {\bf B385} (1996) 125.}

\lref\bds{T. Banks, M. Douglas and N. Seiberg, {\it Probing F-Theory with
branes}, hep-th/9505199.}

\lref\senII{A. Sen, {\it A Non-perturbative Description of the 
Gimon-Polchinski Orientifold}, hep-th/9611186.}

\lref\sag{A. Sagnotti, {\it Open Strings and their Symmetry Groups}, Talk
at Cargese Summer Institute, 1987.}

\lref\prasa{G. Pradisi and A. Sagnotti, Phys. Lett. {\bf B216} (1989) 59.}

\lref\bpsag{M. Bianchi, G. Pradisi and A. Sagnotti, Nucl. Phys. {\bf B376} 
(1992) 365.}

\lref\hor{P. Horava, Nucl. Phys. {\bf B327} (1989) 461, Phys. Lett. {bf B231} 
(1989) 251.}

\lref\dlp{J. Dai, R. Leigh and J. Polchinski, Mod. Phys. Lett. {\bf A4} 
(1989) 2073.}

\lref\lei{R. Leigh, Mod. Phys. Lett. {\bf A4} (1989) 2767,}

\lref\pol{J. Polchinski, Phys. Rev. {\bf D50} (1994) 6041.}

\lref\gimpo{E. Gimon and J. Polchinski, Phys. Rev. {\bf D54} (1996) 1667.}

\lref\dpol{J. Polchinski, Phys. Rev. Lett. {\bf 75} (1995) 4724.}

\lref\bluzaf{J. Blum and A. Zaffaroni, hep-th/9607019.}

\lref\dapark{A. Dabholkar and J. Park, hep-th/9607041.}

\lref\gomu{R. Gopakumar and S. Mukhi, hep-th/9607057.}

\lref\argdo{P. Argyres and M. Douglas, hep-th/9505062.}

\lref\sw{N. Seiberg and E. Witten, Nucl. Phys. {\bf B431} (1994) 484.}

\newsec{Introduction}
Exotic compactifications of type IIB string theory where the ten dimensional
$SL(2,Z)$ modulus of this theory varies over the compact base has received
a lot of attention since the discovery of the F-theory\refs{\vafaI}. In 
\refs{\vafaI} it was shown that F-theory compactified on an elliptically
fibred K3 is equivalent to type IIB string theory on $P^1$ with 24
seven branes, carrying $(p,q)$ charge with respect to the
$SL(2,Z)$ modulus $\tau$. A new understanding of this theory was obtained
when Sen\refs{\senI} showed that in some limit K3 compactification of F-theory
has the coupling, i.e., the $SL(2,Z)$ modulus constant over the base. In this
limit K3 degenerates to $T^4/Z_2$ and the base manifold becomes $T^2/Z_2$. 
Since the coupling is constant and can be tuned to small values using the
tunable parameter, it can be explicitly mapped onto a conventional orientifold 
of type IIB string theory\refs{\sag,\prasa,\bpsag,\hor,\dlp,\lei,\pol}. It was 
shown in \refs{\senI} that the non-perturbative physics of this orientifold is 
described very well by the F-theory and that the mathematically this problem 
is identical to $N=2$ supersymmetric $SU(2)$ Yang-Mills theory coupled to four 
hypermultiplets in the fundamental representation. Using three brane as a 
probe, Banks et al.\refs{\bds} showed how the Seiberg-Witten\refs{\sw} results 
for $N=2$ Supersymmetric Yang-Mills are related to this F-theory 
compactification. Dasgupta and Mukhi\refs{\mukhi} studied enhanced exceptional 
gauge symmetries at certain points in the moduli space of K3 compactification 
of F-theory. More importantly they obtained this enhanced gauge symmetry at a 
coupling which is constant but of the order of 1. Therefore these are 
non-perturbative enhanced gauge symmetries\refs{\mukhi}. 

Six dimensional compactifications of type IIB theory in the F-theory picture 
was studied by Morrison and Vafa\refs{\vafaII}. They studied various models 
using elliptically fibred Calabi-Yau manifolds. Relation of these 
compactifications to the orientifolds of type IIB string theory has been
studied by several authors\refs{\bluzaf,\dapark,\gomu}. Another class of six
dimensional orientifold models were proposed by Gimon and Polchinski
\refs{\gimpo}. These models have $N=1$ supersymmetry in six dimensions and
consist of orientifold five plane and orientifold nine planes along with
Dirichlet five branes and Dirichlet nine branes\refs{\dpol}. Non-perturbative 
physics of the T-dual version of this model was studied recently by Sen
\refs{\senII}. 

In this paper we will study non-perturbative enhanced gauge symmetries in the
T-dual version of the Gimon-Polchinski model. This is analogous to the work
of Dasgupta and Mukhi\refs{\mukhi} except for the fact that there is an
addtional twisting built into the Gimon-Polchinski model which eventually
gives rise to different gauge symmetries. In section 2 we will briefly
discuss results of Sen\refs{\senI} and of Dasgupta and Mukhi\refs{\mukhi}.
In section 3, after briefly introducing the Gimon-Polchinski model and its
T-dual version, we discuss non-perturbative enhanced gauge symmetry in this
T-dual model.

\newsec{Orientifold and F-theory on elliptic K3} 
The simplest example of orientifold is when type IIB theory is compactified 
$T^2/Z_2$, which is related to the F-theory compactification on an 
elliptically fibred K3 surface\refs{\senI}.
Compactifications of the type IIB string theory in which the SL(2,Z)
modulus of the ten dimensional type IIB string theory is allowed to vary 
over a compact base manifold were originally referred to as F-theory. The 
$SL(2,Z)$ modulus of the type IIB theory is identified with the modular
parameter of the elliptic fibre\refs{\vafaI}. The $T^2/Z_2$ orientifold of the 
type IIB string theory corresponds to a certain limit in the moduli space of 
K3 compactifications of the F-theory where the SL(2,Z) modulus of the type IIB 
is constant over the base. In this limit the elliptic fibre has a constant 
complex structure modulus $\tau$ except at four fixed points of the $Z_2$ 
action on the torus where the fibre degenerates. This particular limit 
corresponds to K3 degenerating to $T^4/Z_2$\refs{\senI}.

The Weierstrass equation for elliptically fibred K3 can be written as
\eqn\kthree{y^2=x^3+f(z)x+g(z),}
where $z$ is a coordinate on the base $P^1$ and $f$ and $g$ are polynomials
in $z$ of degree 8 and 12 respectively. The modular parameter $\tau$ of the 
fibre is given in terms of the $j$-function, which, in terms of the
polynomials $f$ and $g$, is given by
\eqn\jfun{j(\tau)={4(24f)^3\over 4f^3+27g^2}.}
The discriminant is given by
\eqn\discri{\Delta(z)=4f(z)^3+27g(z)^2,}
zeroes of which, in F-theory picture, correspond to locations of 
a seven branes each producing deficit angle of $\pi/6$. The 
F-theory picture is related to the orientifold description when the modular 
parameter of the fibre is constant over the base. This corresponds to taking 
$f^3/g^2$ equal to a constant. The simplest choice which allows us to set 
$f^3/g^2$ equal to a constant is
\eqn\fng{f(z)=\alpha\phi^2(z),\quad g(z)=\phi^3(z),}
where $\phi(z)$ is a polynomial in $z$ of degree 4 and $\alpha$ is a relative
scale parameter which can be adjusted to make $Im(\tau)$ large. This not only
reduces the F-theory compactification to a conventional string theory 
compactification but also allows us to use perturbative string methods by 
suitably adjusting $\alpha$.

The choice \fng\ reduces the base to $T^2/Z_2$ orbifold and at the same time 
the fibre $T^2$ develops an $SL(2,Z)$ monodromy
\eqn\mon{\pmatrix{-1&0\cr 0&-1}}
around the all the four orbifold singularities. This orbifold with the 
$SL(2,Z)$ monodromy \mon\ is an orientifold 
$T^2/{\cal I}_z\cdot(-1)^{F_L}\cdot\Omega$ of the type IIB string theory
\refs{\senI}, where ${\cal I}_z$ changes the sign of both the torus 
coordinates, $(-1)^{F_L}$ changes the sign of all the Ramond sector states
on the left and $\Omega$ denotes world-sheet parity transformation. The 
polynomial $\phi(z)$ in this case is given by
\eqn\pphi{\phi(z)=\prod_{i=1}^4(z-z_i)\Rightarrow 
\Delta =(4\alpha^3+27)\prod_{i=1}^4(z-z_i)^6}
where $z_i$, $i=1,..4$ are the locations of the orientifold planes. Analysing
the singularity of the Weierstrass equation for elliptically fibred K3 in the 
neighbourhood of $z_i$ shows that the singularity is of type $D_4$. Thus the
orientifold limit of the F-theory compactification on an elliptically fibred
K3 gives rise to the gauge symmetry $SO(8)^4$. In the orientifold picture 
this gauge symmetry is a consequence of putting four coincident seven branes
on each orientifold plane. We can move individual seven branes away from the 
orientifold plane. This gives rise to non-perturbative physics of this 
orientifold which is nicely captured by F-theory. This F-theory description 
and its connection with the Seiberg-Witten theory is explored in detail in a 
beautiful paper by Sen\refs{\senI}. Interpretation of these results in terms 
of three brane probe is given in \refs{\bds}.

Apart from this choice of $f$ and $g$ there exist other subspaces in the 
F-theory moduli space where we can get constant coupling\refs{\mukhi}. These 
points correspond to either $f(z)=0$ or $g(z)=0$. The $f=0$ case corresponds 
to $j(\tau)=0$ which in turn implies $\tau=\exp(i\pi/3)$. On the other hand, 
$g=0$ corresponds to $j(\tau)=13824$ and $\tau=i$. In order to interpret
these compactifications as conventional compactifications we look for points
in these subspaces which correspond to orbifold singularity. The deficit angle 
$\Delta\theta$ at the orbifold singularity is given by
\eqn\deficit{\Delta\theta=2\pi(1-{1\over n}).}

If $f(z)=0$, then there exist two point in the moduli space where we get 
conventional orbifold singularities along with non-trivial monodromies around
the singular points. These orbifold singularities give rise to enhanced gauge 
symmetries. A point in this moduli space where 
\eqn\esixg{g(z)=(z-z_1)^4(z-z_2)^4(z-z_3)^4,}
and consequently,
\eqn\esix{\Delta=(z-z_1)^8(z-z_2)^8(z-z_3)^8}
gives the enhanced gauge symmetry $E_6\times E_6\times E_6$. The deficit
angles at the locations of 8 seven branes is $4\pi/3$. At this point K3 
degenerates to $T^4/Z_3$ orbifold.

The other point on this branch where we get orbifold singularity is given by
\eqn\eateg{g(z)=(z-z_1)^5(z-z_2)^4(z-z_3)^3,}
which corresponds to
\eqn\eeight{\Delta=(z-z_1)^{10}(z-z_2)^8(z-z_3)^6.}
The enhanced gauge symmetry at this point is $E_8\times E_6\times SO(8)$.
The deficit angles at the locations of 10, 8 and 6 seven branes is $5\pi/3$,
$4\pi/3$ and $\pi$ respectively. The K3 surface with this type of singularity 
structure is $T^4/Z_6$. 

In both the cases discussed above seven branes can move only in pairs. This
comes from the fact that $f(z)=0$ and $\Delta(z)\sim g(z)^2$. Therefore at
every location of the singularity we have atleast a double zero. Let us now
consider $g(z)=0$. On this branch there is only one point where we get 
conventional orbifold singularity. The polynomial $f(z)$ at this point is
given by
\eqn\esevenf{f(z)=(z-z_1)^3(z-z_2)^3(z-z_3)^2,}
and
\eqn\eseven{\Delta=(z-z_1)^9(z-z_2)^9(z-z_3)^6.}
This branch has $g(z)=0$ and $\Delta\sim f(z)^3$. Therefore, $\Delta$ has 
zeroes in multiples of three and, hence, on this branch only 3 seven branes 
can move together. Deficit angles produced at the locations of 9 and 6 seven 
branes is $3\pi/2$ and $\pi$ respectively. Due to this the K3 surface at this 
point degenerates to $T^4/Z_4$.

\newsec{Enhanced Gauge symmetries in Gimon-Polchinski Model}
The Gimon-Polchinski orientifold is obtained by studying type IIB string 
theory on $T^4/(Z_2\times Z_2)$, where one of the $Z_2$ is the world-sheet
parity transformation $\Omega$ and other $Z_2$ changes sign of all four torus 
coordinates $(x^6,x^7,x^8,x^9)$ and simultaneously reverses the orientation of
the world-sheet by parity transformation $\Omega$. The T-dual version of the 
Gimon-Polchinski orientifold and its non-perturbative description was studied 
by sen\refs{\senII}. The Gimon-Polchinski orientifold contains orientifold five
planes and nine planes along with Dirichlet five branes and nine branes. If
we perform T-duality transformation on two of the torus coordinates, say $x^6$
and $x^7$, the orientifold nine plane and Dirichlet nine branes transform into 
orientifold seven plane and Dirichlet seven branes transverse to $x^6-x^7$
plane. The orientifold five plane and Dirichlet five branes on the other hand
transform into orientifold seven plane and Dirichlet seven branes which are
transverse to $x^8-x^9$ plane. The two $Z_2$ transformations after T-duality 
are transformed into
\eqn\ztwo{g={\cal I}_{67}\cdot(-1)^{F_L}\cdot\Omega,\quad {\rm and} \quad
h={\cal I}_{89}\cdot(-1)^{F_L}\cdot\Omega,}
where, ${\cal I}_{67}$ changes the sign of $x^6$ and $x^7$ whereas 
${\cal I}_{89}$ changes the sign of $x^8$ and $x^9$. We choose the complex 
coordinates on $T^4$ such that
\eqn\comp{w=x^6+ix^7,\qquad\quad z=x^8+ix^9.}
These coordinates change sign under $g$ and $h$ respectively. The coordinates
which do not transform under $g$ and $h$ are given by $u=w^2$ and $v=z^2$.
This theory with intersecting orientifolds naively looks like a product of
two orientifold theories studied in the previous section and we would therefore
expect the gauge group, with constant coupling over the base with both
$f$ and $g$ non-zero, to be $SO(8)^4_u\times SO(8)^4_v$. However, projection
by $h$ acts as a discrete gauge transformation on the orientifold generated by
$g$ and vice versa. This discrete transformation is given by
\eqn\discrete{{\cal M} = \pmatrix{0&I_4\cr -I_4&0},}
where $I_4$ is $4\times 4$ identity matrix. The subgroup of $SO(8)$ that 
commutes with this transformation is $U(4)$. The $SO(8)$ adjoint 
hypermultiplet picks up a minus sign when conjugated by ${\cal M}$. It 
therefore gives two hypermultiplets in the {\bf 6} representation of $SU(4)$.
Putting four coincident seven branes on the orientifold plane in this case
gives rise to $U(4)$ gauge symmetry. Looking at the action of ${\cal M}$ on the
adjoint hypermultiplet it is clear that the seven branes cannot move 
individually but can move only in pairs. Thus the total gauge symmetry is
$U(4)^4_u\times U(4)^4_v$. Since this theory involves intersecting seven 
branes, there exist open strings starting from seven branes parallel to $u$ 
plane and ending on seven branes parallel to $v$ plane. These open strings 
correspond to ({\bf 4},{\bf 4}) representation of the $U(4)_u\times U(4)_v$.
Non-perturbative description of this orientifold corresponds to moving the
seven branes pairwise away from the orientifold planes. This is discussed in 
great detail in \refs{\senII}.

Elliptically fibred Calabi-Yau threefold can be defined by
\eqn\calabi{y^2=x^3+f(u,v)x+g(u,v),}
wherer $u$ and $v$ are coordinates on the base $P^1\times P^1$, and $f$, $g$
are polynomials of degree 8 and 12 respectively in both variables $u$ and $v$.
The modular parameter $\tau$ of the fibre is given in terms of the 
$j$-function, which is given by
\eqn\cyj{j(\tau)={4(24f)^3\over 4f^3+27g^2}.}
The discriminant is again given by
\eqn\discricy{\Delta(u,v)=4f(u,v)^3+27g(u,v)^2.}
Every zero of the discriminant locus indicates location of a seven brane.
At the $U(4)^4_u\times U(4)^4_v$ point, the $SL(2,Z)$ modulus of type IIB
theory does not vary over the base, i.e., the modular parameter $\tau$ of the 
elliptic fibre is constant except at the locations of the orientifold planes. 
Constant $\tau$ implies $f(u,v)^3/g(u,v)^2$ is constant. In this limit we can 
write $f(u,v)=f_1(u)f_2(v)$ and $g(u,v)=g_1(u)g_2(v)$, and both $f_1(u)$, 
$g_1(u)$ and $f_2(v)$, $g_2(v)$ are given by eq.\fng\ .

There exist other branches of the ``moduli space'' of Gimon-Polchinski 
orientifold where we can get constant coupling. In these cases too we can
write $f$ as a product of $f_1$ and $f_2$ and $g$ as a product of $g_1$ and
$g_2$. These points correspond to either $f(u,v)=0$
or $g(u,v)=0$. Again $f=0$ case corresponds to $\tau=\exp(i\pi/3)$ and $g=0$ 
corresponds to $\tau=i$. Note that $f(u,v)=0$ can be arranged by either 
setting $f_1$ or $f_2$ to zero or both. Similarly $g(u,v)=0$ can be arranged 
by either setting $g_1$ or $g_2$ to zero or both. This gives rise to several 
combinations. 

Let us consider the case when both $f_1(u)$ and $f_2(v)$ are zero. This gives 
rise to four possibilities. In all these cases $\Delta\sim g^2$ and therefore 
$\Delta$ has zeroes in multiples of two. This means seven branes in these 
cases can only be moved in pairs. 

In the first case the discriminant is given by
\eqn\ufive{\Delta=(u-u_1)^8(u-u_2)^8(u-u_3)^8(v-v_1)^8(v-v_2)^8(v-v_3)^8.}
This configuration has three bunches of 8 seven branes each in $u$ as well as
in $v$ plane. As mentioned earlier every seven brane contributes the deficit 
angle $\pi/6$. Therefore, a bunch of 8 seven branes gives rise to a deficit
angle of $4\pi/3$ which is equal to the deficit angle produced by a $Z_3$
orbifold. In case of \ufive , we see that both $u$ and $v$ planes reduce to
$Z_3$ orbifolds and the elliptically fibred Calabi-Yau threefold becomes a
$T^6/Z_3\times Z_3$ orbifold. Looking at the singularities of \ufive\ we would
naively expect the gauge symmetry generated by these singularities to be 
$(E_6\times E_6\times E_6)_u\times(E_6\times E_6\times E_6)_v$.
In our case, however, there is an additional twist generated by the discrete
gauge transformations. In the generic constant coupling case, i.e., both $f$
and $g$ non-zero, these discrete gauge transformations broke the gauge group
$SO(8)^4_u\times SO(8)^4_v$ down to $U(4)^4_u\times U(4)^4_v$. To determine 
which subgroup of $E_6$ remains unbroken after twisting, all we need to do
is to embed $SO(8)$ in the subgroup of $E_6$ and study the action of the 
discrete gauge transformation \discrete\ on this subgroup. This twisting will 
break the subgroup of $E_6$ to a smaller group which commutes with the 
discrete gauge transformation. The consistency condition, of course, is that 
when this subgroup of $E_6$ breaks to $SO(8)$, the corresponding group after 
twisting should reduce to $U(4)$. We can embed $SO(8)$ in the $SO(10)$ 
subgroup of $E_6$. The discrete gauge transformation breaks $SO(10)$ to $U(5)$ 
in such a way that when $SO(10)$ breaks to $SO(8)$, $U(5)$ breaks to $U(4)$. 
Thus, we find that total gauge symmetry which obeys the consistency condition 
is $(U(5)\times U(5)\times U(5))_u\times (U(5)\times U(5)\times U(5))_v$. 
Twisting acts on the adjoint hypermultiplet of $SO(10)$ to give two 
hypermultiplets in the {\bf 10} representation of each $SU(5)$.

In the second case the discriminant is given by
\eqn\ueight{\Delta=(u-u_1)^{10}(u-u_2)^8(u-u_3)^6(v-v_1)^{10}(v-v_2)^8
(v-v_3)^6.}
Here the seven branes are bunched together both on $u$ and $v$ planes into
three sets of 10,8 and 6 seven branes. They give the deficit angle of 
$5\pi/3$, $4\pi/3$ and $\pi$ respectively. Since the seven branes are bunched 
together in the same fashion on both $u$ and $v$ planes, the elliptically
fibred Calabi-Yau threefold in this case degenerates to a $T^6/Z_6\times Z_6$
orbifold. If we ignore the twisting for the time being then Tate's algorithm
predicts that the gauge symmetry is $(E_8\times E_6\times SO(8))_u\times
(E_8\times E_6\times SO(8))_v$. Now let us consider the action of the discrete
gauge transformation. We will again use the same method explained above for 
the $E_6$ symmetry. The $E_8$ gauge symmetry after twisting breaks down to
$U(8)$. In this case we embed $SO(8)$ in the $SO(16)$ subgroup of $E_8$. 
Total gauge symmetry therefore is $(U(8)\times U(5)\times U(4))_u\times
(U(8)\times U(5)\times U(4))_v$. In case of $E_8$, the adjoint hypermultiplet
reduces to two hypermultiplets in the {\bf 28} representation of each $SU(8)$.
We also have two hypermultiplets in the {\bf 10} representation of each
$SU(5)$ and two hypermultiplets in the {\bf 6} representation of each $SU(4)$.

The discriminant in the third and fourth case are 
\eqn\eightfive{\Delta=(u-u_1)^{10}(u-u_2)^8(u-u_3)^6(v-v_1)^8(v-v_2)^8
(v-v_3)^8,}
and
\eqn\fiveight{\Delta=(u-u_1)^8(u-u_2)^8(u-u_3)^8(v-v_1)^{10}(v-v_2)^8
(v-v_3)^6.}
Notice that these two cases are related to each other under the exchange of
$u$ and $v$. In this case the elliptically fibred Calabi-Yau threefold
degenerates to $T^6/Z_3\times Z_6$. The gauge symmetry after twisting is 
$(U(8)\times U(5)\times U(4))_u\times(U(5)\times U(5)\times U(5))_v$, and
similarly for the fourth case by switching $u$ and $v$. Again we have two 
hypermultiplets in the representation {\bf 10} of each $SU(5)$, two in the 
representation {\bf 6} of each $SU(4)$ and two in representation {\bf 28} of 
each $SU(8)$.

Now let us consider the case when $g(u,v)=0$ and $f$ nonzero. We will first 
consider the case when both $g_1(u)$ and $g_2(v)$ are zero. In this case 
$\Delta\sim f^3$ so whenever $\Delta$ vanishes it has atleast a third order 
zero or multiples of 3. That means on this branch seven branes can be moved 
around in the bunches of three. There is only one possibility in this case and 
the discriminant is given by
\eqn\usix{\Delta=(u-u_1)^9(u-u_2)^9(u-u_3)^6(v-v_1)^9(v-v_2)^9(v-v_3)^6.}
Both on $u$ and $v$ plane the seven branes are bunched together into three
sets of 9,9 and 6 seven branes. The deficit angles due to these coincident
seven branes is $3\pi/2$, $3\pi/2$ and $\pi$ respectively. Comparing with the
orbifold deficit angles we see that the elliptically fibred Calabi-Yau 
threefold has degenerated into $T^6/Z_4\times Z_4$ orbifold. If we ignore
twisting then the expected gauge group is
$(E_7\times E_7\times SO(8))_u\times(E_7\times E_7\times SO(8))_v$. After 
including the effect of twisting this gauge group breaks to 
$(U(6)\times U(6)\times U(4))_u\times(U(6)\times U(6)\times U(4))_v$. The 
subgroup of $E_7$ which contains $SO(8)$ is $SO(12)$, which breaks to $U(6)$
due to the action of the discrete gauge transformation. We get two 
hypermultiplets in the representation {\bf 15} of each $SU(6)$ and
two hypermultiplets in the representation {\bf 6} of each $SU(4)$.

The cases we studied so far have either $f_1(u)$ and $f_2(v)$ equal to zero or
$g_1(u)$ and $g_2(v)$ equal to zero. As mentioned earlier, in case of constant
$\tau$ or equivalently constant $SL(2,Z)$ modulus, $f$ can be written as a
product of $f_1$ and $f_2$ and similarly $g$ can be written as a product of
$g_1$ and $g_2$. Now we will consider cases when $f_1=0$ but not $f_2$
and similarly cases when $g_1=0$ but not $g_2$. Other possibilities can be 
obtained by simply swapping $u$ and $v$. Note that all these configurations
can be obtained by deforming either the configurations $f_1=f_2=0$ or
$g_1=g_2=0$. We would also like to note at this point that most of these
possibilities do not correspond to any conventional compactification of type 
IIB string theory. This is because as seven branes move away from each other 
the deficit angles that we get from the new configurations fail to satisfy the 
orbifold deficit angle condition, i.e., $\Delta\theta = 2\pi(1-1/n)$.

Let us consider the case $f_1(u)=0$ and no restriction on $f_2(v)$. Since 
$f(u,v)=f_1(u)f_2(v)$, we see that the $j$-function vanishes no matter what 
$f_2(v)$ is and hence the modular parameter $\tau=\exp(i\pi/3)$. Since there is
no restriction on the form of the polynomial $f_2(v)$ and $g_2(v)$, we will
take the most general choice of these functions consistent with the 
constraints. This choice is given by
\eqn\seiwi{\eqalign{f_2(v)&=\prod_{i=1}^4 f_{SW}(v;m_{1i},m_{2i},m_{1i},m_{2i},
\tau_0)\cr 
g_2(v)&=\prod_{i=1}^4 g_{SW}(v;m_{1i},m_{2i},m_{1i},m_{2i},\tau_0),}}
where, $f_{SW}$ and $g_{SW}$ are the functions that appear in the 
Seiberg-Witten curve for the N=2 supersymmetric $SU(2)$ gauge theory with
four hypermultiplets in the fundamental representation. It is known that
this configuration in the $v$ plane breaks the gauge symmetry\refs{\senII} to 
$(SU(2)_v\times SU(2)'_v)^4$. Now we will turn our attention to the function
$g_1(u)$. Orbifold singularity in the $u$ plane can be obtained for two
choices of the function $g_1(u)$. The first choice is
\eqn\gndel{g_1(u)=\prod_{i=1}^{3}(u-u_i)^4.}
The discriminant that we get from this is given by
\eqn\swdis{\Delta\sim\prod_{i=1}^3(u-u_i)^8 
\prod_{j=1}^4 g_{SW}(v;m_{1j},m_{2j},m_{1j},m_{2j},\tau_0)^2.}
This orbifold singularity in the $u$ plane gives the gauge group $(U(5)\times
U(5)\times U(5))_u$. Thus the total gauge symmetry is 
$(U(5)\times U(5)\times U(5))_u\times(SU(2)_v\times SU(2)'_v)^4.$
In the limit $m_{1k}=m_{2k}$ for a particular value of $1\leq k\leq 4$ then 
the gauge symmetry $SU(2)_{vk}\times SU(2)'_{vk}$ gets enhanced to 
$Sp(4)_{vk}$. Gauge group $U(4)_{vk}$ is recovered when $m_{1k}=m_{2k}=0$.
A branch of the moduli space where $m_{1k}=m_{2k}=0$ for all $k$ corresponds
to  elliptically fibred Calabi-Yau threefold degenerating to 
$T^6/Z_3\times Z_2$ orbifold. 

Second choice for the function $g_1(u)$ is given by
\eqn\sechoice{g_1(u)=(u-u_1)^5(u-u_2)^4(u-u_3)^3.}
The discriminant $\Delta(u,v)$ in this case is given by
\eqn\secdel{(u-u_1)^{10}(u-u_2)^8(u-u_3)^6\prod_{j=1}^4
g_{SW}(v;m_{1j},m_{2j},m_{1j},m_{2j},\tau_0)^2.}
The orbifold singularity in $u$ plane gives the gauge group
$(U(8)\times U(5)\times U(4))_u$ and hence the total gauge symmetry is
$(U(8)\times U(5)\times U(4))_u\times(SU(2)_v\times SU(2)'_v)^4$. As mentioned
earlier at special points $SU(2)_v\times SU(2)'_v$ gauge symmetry is enhanced
to either $Sp(4)_v$ or $U(4)_v$. In this case there exists a branch of the 
moduli space where elliptically fibred Calabi-Yau threefold degenerates to
$T^6/Z_6\times Z_2$ orbifold.

Let us consider the branch on which $g_1(u)=0$ and there is no restriction
on $g_2(v)$. The $j$-function in this case is equal to 13824 and $\tau=i$.
Again we can choose most general form for $f_2(v)$ and $g_2(v)$ which is
given by the Seiberg-Witten ansatz $F_{SW}$ and $g_{SW}$. The choice of 
$f_1(u)$, and hence the enhanced gauge symmetry, is given by the orbifold 
singularity in the $u$ plane, 
\eqn\usixf{f_1(u)=(u-u_1)^3(u-u_2)^3(u-u_3)^2.}
The discriminant is given by
\eqn\usidel{\Delta(u,v)= (u-u_1)^9(u-u_2)^9(u-u_3)^6\prod_{j=1}^4
f_{SW}(v;m_{1j},m_{2j},m_{1j},m_{2j},\tau_0)^3.}
The gauge symmetry in this case is given by $(U(6)\times U(6)\times U(4))_u
\times (SU(2)_v\times SU(2)'_v)^4$. As usual at special points $SU(2)_v\times 
SU(2)'_v$ gauge symmetry is enhanced to either $Sp(4)_v$ or $U(4)_v$. Here 
elliptically fibred Calabi-Yau threefold degenerates to $T^6/Z_4\times Z_2$
orbifold when all the mass parameters are set to zero.

We will now go back to the models where either both $f_1(u)$ and $f_2(v)$ were
zero or both $g_1(u)$ and $g_2(v)$ are zero. Since these models contain 
intersecting orientifolds and intersecting seven branes, there exist 
additional massless states coming from open strings starting from seven branes 
parallel to $u$ plane and ending on seven branes parallel to $v$ plane. Let us 
now consider $f(u,v)=0$ in such a way that both $f_1(u)$ and $f_2(v)$ are
zero. The gauge symmetry has the form $(U(n_1)\times U(n_2)
\times U(n_3))_u\times(U(n_4)\times U(n_5)\times U(n_6))_v$ where $n_i=4,5,8$ 
for $0\leq i\leq 7$. The hypermultiplets that we get from the open strings 
stretching between the intersecting seven branes belong to 
$({\bf n}_j,{\bf n}_k)$ representation where $j=1,2,3$ and $k=4,5,6$. Let us 
consider the configuration of open strings for $U(5)_u\times U(5)_v$ gauge 
group. The hypermultiplet that we get from the open strings stretching between 
these configurations belongs to $({\bf 5},{\bf 5})$ representation. In case of 
$U(8)_u\times U(8)_v$ the hypermultiplet is in $({\bf 8},{\bf 8})$ 
representation. Recall that we get $U(5)_u\times U(5)_v$ gauge symmetry when
$\Delta\sim (u-u_1)^8(v-v_1)^8$. Let us compare this with $U(4)_u\times U(4)_v$
gauge symmetry where $\Delta\sim (u-u_1)^6(v-v_1)^6$. Hypermultiplet in
$({\bf 4},{\bf 4})$ representation as well as the $U(4)_u\times U(4)_v$ 
implies existance of four seven branes at $u_1$ in $u$ plane and four seven
branes at $v_1$ in $v$ plane. But $\Delta$ has a sixth order zero at those
points when we expect only fourth order zeroes there. The interpretation of
the additional double zero\refs{\senI,\senII} is that the non-perturbative
effects split the orientifold plane located there into two seven branes.
Let us go back to $U(5)_u\times U(5)_v$. In this case $\Delta$ has eighth
order zero at $u=u_1$ and $v=v_1$. The gauge symmetry and the open string
hypermultiplets imply that there are only five seven branes. The interpretation
of the additional triple zero of the $\Delta$ therefore is that the
generalised orientifold\refs{\mukhi} in this case splits into 3 seven branes
as we deform the configuration. On the other hand, in case of
$U(8)_u\times U(8)_v$ symmetry $\Delta\sim (u-u_1)^{10}(v-v_1)^{10}$ has 
tenth order zero at $u=u_1$ and $v=v_1$. Hence if we deform this 
configuration, the relevant generalised orientifold splits into only a pair 
of seven branes.

We will now turn to the branch where $g(u,v)=0$ as well as $g_1(u)=0$ and
$g_2(v)=0$. The gauge symmetry at the special point is 
$(U(6)\times U(6)\times U(4))_u\times (U(6)\times U(6)\times U(4))_v$.
The hypermultiplets belonging to every combination $U(n)_u\times U(m)_v$
are in representation $({\bf n},{\bf m})$, where $n,m=4,6$. Let us again
consider a specific situation where we have $U(6)_u\times U(6)_v$ gauge 
symmetry. We get this gauge symmetry when $\Delta\sim (u-u_1)^9(v-v_1)^9$.
We again see that both hypermultiplets and the gauge symmetry suggest
existance of 6 seven branes at those points. The additional triple zero
of $\Delta$ can be interpreted as the generalised orientifold splitting
into 3 seven branes as we deform the configuration. Here we would like to 
point out that we can see the effect of generalised $U(6)$ orientifold 
splitting into 3 seven branes by starting from the perturbative vacuum and 
going backwards to construct the nineth order zero of $\Delta$ by deforming 
the seven brane configuration. To see this let us start with the most general 
configuration in, say, $u$-plane. Here we will concentrate only on the 
$u$-plane and the existance of $v$-plane is felt only through the twisting 
constraint. Most general configuration in $u$-plane corresponds to 
$(SU(2)\times SU(2)')^4$ gauge symmetry. Two $SU(2)$ factors come from two 
pairs of seven branes and fourth power occurs because we have four sets of 4 
seven branes located originally at four orientifold planes. As the seven 
branes move away pairwise the gauge symmetry breaks to $SU(2)\times SU(2)'$ 
near every orientifold. Another thing that happens due to this is the 
orientifold plane splits into 2 seven branes. Unlike the other seven branes 
which carry elementary seven brane charge these seven branes in general carry 
dyonic charge\foot{ By dyonic charge we mean the seven brane certainly 
carries magnetic charge. Some of them have no fundamental seven brane charge 
whereas some others do.} and do not 
move in pair. So the most general configuration has eight pairs of seven 
branes and eight dyonic seven branes which are not paired. By taking one
of the seven brane pairs on the top of one of the dyonic seven brane we
can obtain an Argyres-Douglas type point\refs{\argdo}. The most symmetric 
configuration corresponds to every dyonic seven brane capturing one paired
seven brane each. This gives us eight points of Argyres-Douglas type, i.e.,
every point has two seven branes with fundamental charge and one with a
dyonic charge. Now $(U(6)\times U(6)\times U(4))_u$ gauge group comes from
the configuration where we club eight Argyres-Douglas points into three sets
(3,3,2). Confluence of three Argyres-Douglas points gives rise to $U(6)$
symmetry. Now we can easily see that at this point we have six fundamental
seven branes which is consistent with the gauge symmetry as well as with
the open string hypermultiplets and we also have three dyonic seven branes
which compensate for the additional zeroes occuring in the discriminant
$\Delta$. 

It would be nice to have similar understanding of the generalised
orientifold which gives $U(5)$ gauge symmetry. So far we have not succeeded
in extracting that information. We hope our results will help shed more light 
on the structure of generalised orientifolds.

\bigskip
\noindent{\bf Acknowledgements:} We would like to thank A. Sen for patient
explanation of his results and several valuable suggestions and discussions. 
We would also like to thank A. Biswas, A. Kumar and S. K. Rama for useful 
discussions.
\listrefs
\bye